\documentclass[aps,prb,reprint]{revtex4-1}
\usepackage{graphicx}
\usepackage{hyperref}
\newcommand\Fig[1]{Fig.~\ref{#1}}
\begin{document}

\title{Coming home from a MOOC}

\author{Werner Krauth}
\email{Director of the Physics Department at \'Ecole normale sup\'erieure,
werner.krauth@ens.fr}
\affiliation{Laboratoire de Physique Statistique, \'Ecole Normale
Sup\'erieure,
UPMC, Universit\'e Paris Diderot, CNRS, 24 rue Lhomond, 75005 Paris, France}
\maketitle 

\section{The MOOC battle field}

In recent years, Massive Open Online Courses (MOOCs) have been in
the public discussion in Europe, as much as in the United States. I
had kept myself a safe distance from the debate until, about a year
ago, \'Ecole normale sup\'erieure (ENS) asked me to participate
in a pilot project to set up and teach a masters-level MOOC in
statistical/computational physics. The ENS project also included a course
on French philosophy,\cite{philos} and one on Galois theory,\cite{galois}
a branch of mathematics. A recording studio was to be installed on
our campus in the Latin quarter, and videos were to be recorded, cut
and edited by professionals. Courses were to be hosted on the Coursera
platform.\cite{coursera} I would be able to recruit additional teachers
and student assistants. All this effort was quite engaging.

I first checked that my colleagues with whom I had taught a successful
course in our English-language Physics masters program\cite{master} were
interested in adapting the course to the internet.  I then signed up,
not expecting what was going to happen: Our 10-week MOOC on Statistical
Mechanics: Algorithms and Computations\cite{ourcourse}   attracted 30,000
registered students, triggered 256,000 video downloads, and 120,000 visits
to its forum. When it was all over, we were overwhelmed by our teaching
experience with an enthusiastic international and trans-cultural class
of students that neither we nor our university had chosen. We had gone
to the center of core subjects in science, including the nature of phase
transitions, the physics of liquids, random processes, path integrals,
and Bose-Einstein condensation. We had discussed many deep connections
between physics and computing.

By the time it was over, we had presented and prepared for download
about 250 programs (in Python), and initiated a great amount of practical
computing and  program writing and program rewriting in the nine homework
sessions. When we were done, we finally took our eyes off the forum
that for three months had been a buzzing platform of heated discussions
and of mutual help.  We had motivated students toward a common goal,
but had also been boosted by their enthusiasm.

When the course was over, we had not suffered from the absence of a live
audience during lectures, although it is now most rewarding to discuss
with former students when they occasionally come up to us and present
themselves on the street in Paris, New York, or elsewhere at conferences
and seminars.

Students had worked hard, although not for university credit. They had
handed in 6300 homework solutions, and corrected about 20,000 of them
through the peer-correction system that  worked amazingly well. They had
authored 5200 posts on the forum. After the course, 5\% of the students
who had followed the course most diligently indicated that they would use
what they had learned on a daily basis, 23\% several times per month,
and 50\% a few times a year.  At the end of the course, they were as
exhausted as we were. A student from the Netherlands indicated that ``It
was a very challenging experience that totally absorbed me for almost
three months \ldots the coolest thing I ever did \ldots\ ." From Norway,
our own feelings were mirrored: ``Now, the course is over, I look up
from my PC -- the wife and child are still there \ldots\ ." Countless
posts and comments reflected on the experience, the material, and the
medium that it had taken place in.

It is difficult to foresee to what degree MOOCs will permeate core-level
academic training. I am even unable to tell whether the second edition
of our course, in early 2015, will recreate this year's ambiance, and
transmit knowledge, foster communication and exchange as efficiently. The
present text cannot provide a blueprint for online teaching, but it may
enrich the ongoing discussion about MOOCs from the vantage point of a
long-time teacher and recent home-comer from the MOOC battle field.

\section{Teasers, Green screens, Mother of Studies}

\begin{figure}
\includegraphics[width=8cm]{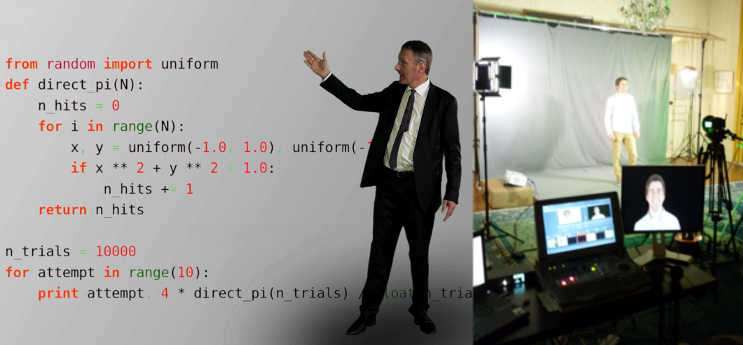}
\caption{Announcement video (``teaser") of our MOOC after professional
editing (left) and during production using a green screen (right).}
\label{figTeaser}
\end{figure}

Our MOOC project came to life four months before the launch date. It
was time to do the promotional video, the only live item to be rendered
visible without registration. Writing this three-minute ``teaser" took
one hour. Recording it brought us into contact with the camera and
editing team from ENS and from FEMIS, the Paris school of cinema. The
video professionals  guided us toward the green screen technology, that
dissociates the recording of the ``actor" from the creation of the
scene: During takes, the otherwise empty scene was bathed in intense
green light that reflected from the background and was then eliminated
(see \Fig{figTeaser}). Text, equations, images,  and animations
were then incorporated by the editor. From the start we were comfortable
with the fact that our pedagogical project was to be realized by videos
that used a genuine video technology (green screen), rather than by the
mere videotaping of a classroom course using classroom tools (such as
a blackboard and PowerPoint).

MOOCs differ from  classroom teaching not only because the medium is
different, but more importantly  because of a difference in viewing modes
and therefore in the relation to time. Students routinely accelerate or slow
down the video player (at constant audio pitch), they stop and  may even quit.
Videos are  not viewed in linear time, and difficult sequences are
replayed over and over again. The power to repeat is now in the hand of
the individual student and it suits individual needs. This power
gives new meaning to the eternal truth that ``repetition is the mother
of studies" and implies that the teachings must be less redundant
than before. Our usual kickoff classroom meeting melted down into
the three-minute teaser, and our 90-minute live classes turned into
half-hour videos, without any loss of content. In the shorter, less
repetitive, videos, even slight imprecisions (such as omitted words and
half sentences) would create terrible confusion, further amplified by the
multiple viewings. To avoid this confusion in our MOOC, the material was
carefully scripted, and all shootings were controlled by a fellow
teacher.  This work flow was initiated with the teaser and  was
kept during the entire course.

In their responses, 30\% of the students participating in the final
survey indicated that the single item that they  liked best during the
course was the quality of the videos. Many messages on the forum and in
the survey made us think that our choices worked well.

\begin{figure}[ht]
\includegraphics[width=6.cm]{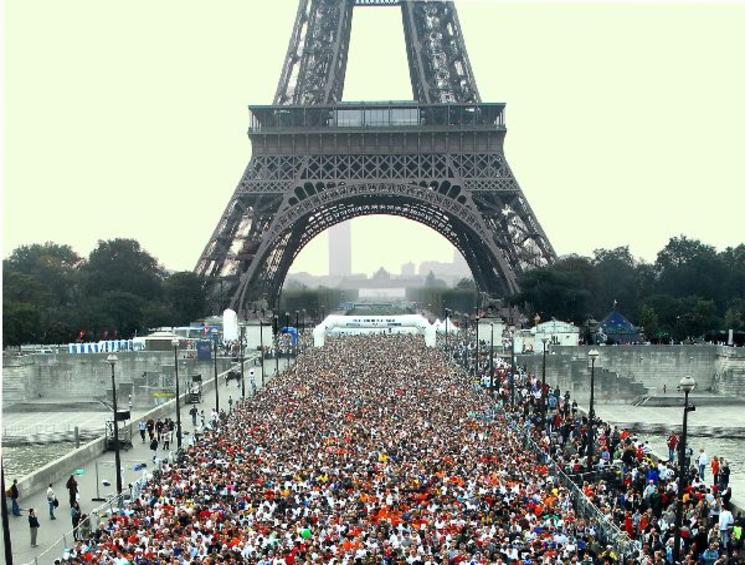}
\caption{Modern marathons can be {\bf M}assive, yet their running
distance is not reduced. Our {\bf M}OOC likewise remained at the level
of an advanced masters course in physics. It appealed to a large number of
people with a variety of personal goals, performances, and levels
of involvement.}
\label{figMarathon}
\end{figure}

\section{Large Numbers, City Marathons}

Registration was free of charge, and there was no academic prerequisite
-- after all, one of the ``O"s in ``MOOC" stands for ``Open." Three
days after posting our teaser on the Coursera website, 667 students had
enrolled. One month later, there were 5000, and this number was still
far from the final count. We came to understand that our MOOC was to
be massive, but sensed that even with thousands of students, we should
continue to  draw on all the advanced mathematical tools of our classes
at ENS, and aim at the same deep physical insights.  We thus confirmed
that, even with a big crowd, we would engage in core-level (rather than
introductory) academic teaching.

Facing massive registrations, we thus arrived at the same conclusion as
the organizers of large-scale sports events. City marathons, for example,
have become extremely popular but they still run over 42.195\,km (see
\Fig{figMarathon}). Like our MOOC, they appeal to a large and quite
ambitious public with different backgrounds and a great span of personal
goals and performance levels.

The large number brings many new opportunities, beyond the fact that it
simply means more people. A larger forum is usually richer, especially
with the modern voting mechanism that allows the best posts to rise to the
top. In fact, MOOC teaching needs a large audience  to ``take off,'' and
certainly more than what a regular classroom and the graduate school of a
single university can offer. Their massive nature is an opportunity, but
also a survival condition to sustain necessary liveliness in teaching. The
sheer size carries its own dangers.  Catastrophic failure, for example the
inability to go through with a course, presents a real threat. Wherever
it happened in other MOOCs, it was widely publicized and discussed.

Among the students participating in the final survey, 20\% had completed
a bachelor's degree (or similar), 40\% had completed a master's degree,
and 30\% a doctorate, and 90\% of them had obtained their degree in a
scientific discipline.  The multiplicity of personal goals was clearly
expressed on the forum. One student had struggled with homework~5,
the first one on quantum mechanics. When he was done, long after
the submission deadline, he opened a thread ``Yay! HW5 finished,"
that drew half a dozen of comments such as: ``Congratulations on
completing HW5!" A number of French post-high school (prep-school) professors
participated in the course, in order to obtain inspiration
for scientific Python programming projects in their own classes.

\section{Going live, Classes}

On February 3, 2014, the first week of our course went online. A lecture,
a tutorial, and a homework session introduced  Markov chains and
Monte Carlo methods. Week two was about the emergence of statistical
mechanics from classical mechanics. In later weeks, we discussed
phase transitions and virial expansions, and derived the Maxwell and
Boltzmann distributions, among many other subjects. Students stayed
tuned when we turned our attention to quantum physics: density matrices,
path integrals, and Bose-Einstein condensation, before returning to
magnetism, optimization methods and L{\'e}vy distributions. Each subject
was illustrated by short Python programs that students could download,
run, and modify. Every week, we opened sub-forums corresponding to the
current themes. Contributions to these sub-forums took off shortly after
the video upload and abated when the corresponding homework assignment
had passed its due date.

Subjects such as the above are usually too difficult to be learned
from books or from videos alone. They are rather mastered by attending
a \emph{class}, abstractly defined as a group of students that meets
regularly during a given time period for instruction on a given subject
(``alumni" status is also often included in the definition). Classes
take place at universities, seminaries, conservatories, etc, and the
organization into classes made them survive the Gutenberg
media revolution in the 15th century, when knowledge became available in
printed books. Classes are also the essence of
present-day MOOCs. One must register (this defines the group) and there
are starting and end dates. Regularity is provided through the weekly
organization. Students meet on the forum and, in our case, during the
mutual correction of homework. The prime originality of MOOCs, compared
to other forms of distance learning, is the consistent organization of
classes with these characteristics.

From the start, students embraced the idea of our MOOC being a class
rather than an illustrated and animated text book.\cite{SMAC} On the first
day, a student spontaneously created a thread named ``Please allow me to
introduce myself \ldots\ ." Its first post, from Norway, drew 69 answers
from fellow students in Canada, the United States, Brazil, Serbia,
Macedonia, Sweden, India, Saudi Arabia, Spain, Argentina, Greece, and
other countries. Students were authorized to create threads. They might
address questions about Python installation on the different platforms,
but also address specific issues with threads such  as ``lecture 5: 11.14
sec" or ``Tutorial 8 $@$ 8:00 characteristic peak \ldots\ ." Initial
posts triggered animated discussions, concerning subjects from computer
issues to Python programming, and to quantum mechanical wave functions
for students who had often never studied the subject.  About 150 posts
on the forum (2.5\% of the total), were written by the teachers.  Many of
them closed threads: they brought an end to a discussion.

There were also many more general threads. One was entitled ``My notes on
the course." It rendered accessible the entire notes from the ten weeks,
nicely written in \LaTeX\ and fully illustrated. It drew 50 ``likes"
and comments as: ``Thanks Tony! Excellent! Top Class notes indeed!!!"
A 15-minute youtube movie\cite{movie} that a student spontaneously created
and then announced on the forum, went over a particularly intricate
six-line algorithm in tutorial~3. It was greeted by a ``Dennis, you are
a hero." Clearly, our MOOC functioned like a class.

\section{Homeworks, Goals, and Grades}
In all but the final week of the course, a homework assignment enhanced
the lecture and the tutorial. The  assignments were progressive, and
the required programming skills increased from week to week. After the
release (typically a four-page text), students had two weeks to hand in
their solutions.  Immediately after the submission deadline, a master
solution was communicated to students who had contributed work, and
detailed instructions for grading were given. Students then anonymously
corrected three (or more) homework solutions of other students. The
entire process was handled on the Coursera website,\cite{coursera} and
it worked flawlessly. Conceiving the assignments and writing the master
solutions required great concentration. The ``grade" was computed from
the points obtained in the homework assignments (50\%) and from a final
two-hour multiple-choice exam (50\%). The overall requirements of our
course were very high, and it was difficult to make the cut to receive
a certificate. The certificate, which was not endorsed by ENS,  did not
provide ``real" university credits. However, reaching the certificate
level constituted an undeniable academic achievement. More than 2000
students stayed with the course until the end, and 350 students earned
the certificate. These 350 students participated in the final survey.

To understand the massive student participation in a course that offered
no degree and few certificates, it is appropriate to analyze students'
motivation. In education research, motivation is understood along two
categories: The first category, the ``learning goal," applies when
students  pursue educational programs (such as lectures and courses)
primarily to increase their competence. In this category, new challenges
are readily accepted, and failure is easily handled. As emphasized in
seminal psychological research\cite{Elliott_Dweck_1988}: ``The focus
of individuals who pursue learning goals (whether they believe their
ability to be high or low) is on improving ability over time, not on
proving current ability. As noted, obstacles will not as readily be
seen to imply goal failure and will, therefore, not require defensive
maneuvers, not as readily generate anxiety, and not detract from the
intrinsic rewards shown to derive from involvement and progress on a
valued task." Clearly,  our MOOC was firmly planted on the learning goal.

In the second category, the ``performance goal," great value is attached
to doing better (in grades) or to avoid doing less well (in grades) than
others.  The performance goal is a standard setting in higher education.
It requires a homogeneous class from which there is a high social price
for dropping out, and great recognition for performing well. However,
the social implications  of the performance goal can lead weaker students
to shy away from challenges, and even the stronger students can be led to
avoid risk.\cite{Elliott_Dweck_1988} Binge studying, self-handicapping,
and cheating have been associated with the performance goal.

In the final survey, 98\% of students who had obtained the certificate
indicated that the absence of a real degree posed  no problem to
them. Students thus confirmed the ``learning goal" motivation. It made
them adventurous and contributed to the excellent atmosphere during
the class. Although students concentrated  on learning, they did voice
their frustrations with the built-in imperfections of the peer-correction
systems. Students were also quite critical of the few delays that occurred
in the uploads of lectures and homework assignments.  Notwithstanding
their learning goals and their great attitude, they perceived themselves
as participating in a class, and they expected the regularity that is
part of its definition.

\section{Party in Pink, Violence of Learning}

\begin{figure}
\includegraphics[width=8cm]{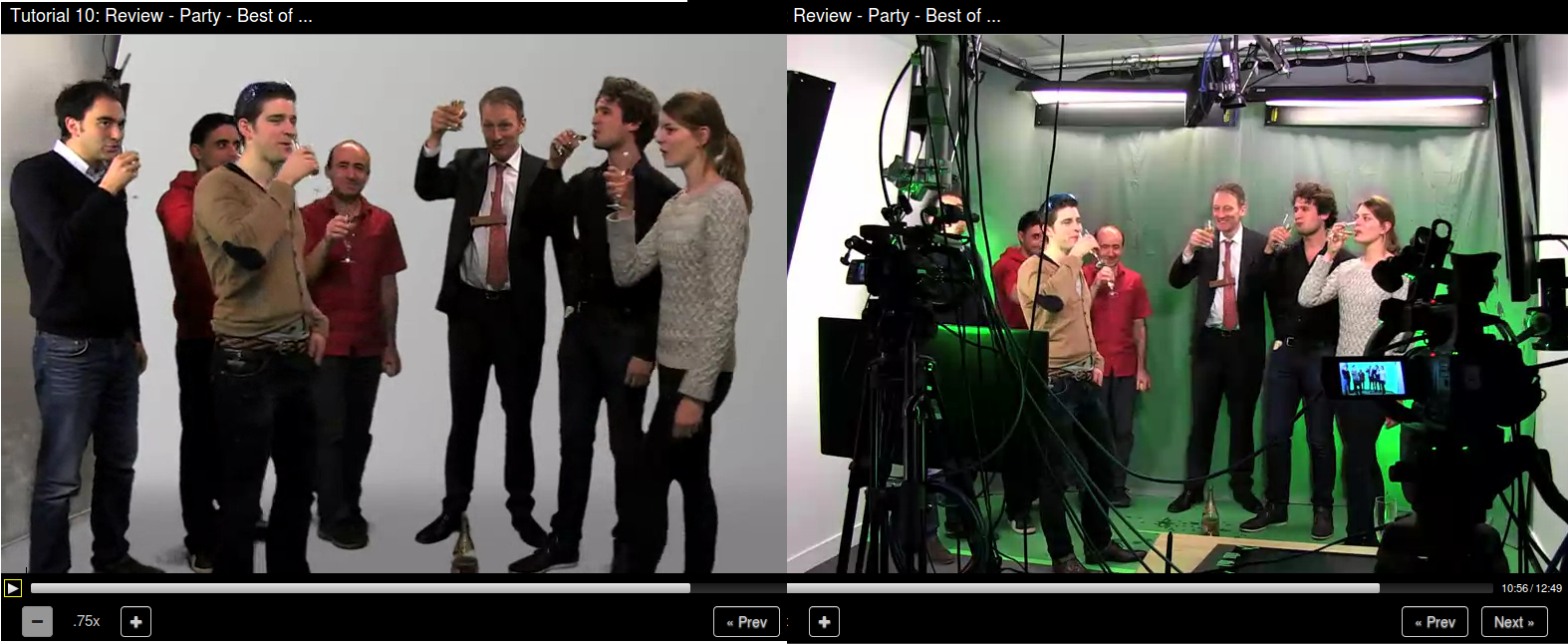}
\caption{Tenth tutorial at 10:56, the final scene of our MOOC, after
professional editing (left) and during production using a green screen (right,
in the studio).} 
\label{figScreenshot_party}
\end{figure}

The course ended, in the final tutorial, with an online party and
with champaign from a pink bottle (because of the green screen, see
\Fig{figScreenshot_party}). Our high-intensity project had been a fierce
battle. With thousands of students watching (and staying), we had been
in a ten-week euphoria. But with hundreds of students reacting week
after week, and scrutinizing every sequence of the videos and struggling
through each sentence of the homework assignments, we were constantly
on alert. Most of the material was created, in anxious euphoria, while
the course was running. Lacking experience, we needed the feedback from
week $n$ to prepare weeks $n+1$ and $n+2$. Fortunately, all the science
had been cleared up well ahead of the launch date and, following expert
advice, we  included no last-minute scientific results. Our course was
a stand-alone project. We did not try  to integrate it into the regular
program at ENS, which would have implied  breaking it up into different
tracks, with proctored exams and staff-graded homework 
for the local students.  The split into different tracks  might have broken the
spell of the course, and destroyed its lighter-than-air atmosphere. Courses such
as ours do not necessarily appeal to all students of physics, but it would be
easy to set up additional exams, outside of the MOOC, to meet requirements for
university credits, at ENS or elsewhere.

It is evident that lecturing  (in MOOCs or in the classroom) is only one
part of our teaching activity, which encompasses lab sessions, practicals, and
small research projects. Teaching also extends to tutoring, the essential
one-to-one relation between a professor and a student, that MOOCs will never
supply. In evaluating MOOCs' possible impact, we must take into account that
learning is a complicated, multifaceted process that consists in the violent
confrontation of the student with the outside world. Even decades after our
student days, we all retain vivid, indelible memories of only a handful of very
special courses spread out over our years of study. Those courses were the ones
that shaped us into what we have become. Time will tell whether the new medium
of MOOCs is strong enough to leave this kind of imprint on students, and whether
tomorrow's professionals will have received formative influence from
courses delivered over the internet. The answer to this question is
not clear to me, but our first attempt to find out was sufficiently
encouraging to try again, in February 2015, for the second edition of
Statistical Mechanics: Algorithms and Computations.

\begin{acknowledgments}
Alberto Rosso, Michael K\"opf and Vivien Lecomte taught in the tutorials
and participated very actively in all stages of our MOOC. Tommaso
Comparin provided final versions of Python programs and produced the
English subtitles. Maxim Berman produced the computer animations inside
the videos.  Emilie Noblet directed the studio. Baptiste Ribrault
was the editor. Additional camera was done by Nordine Meziane and by
Fr\'ed\'eric Borja. Yves Laszlo was in charge of the ENS MOOC program. My
deep gratitude goes to all of them.
\end{acknowledgments}

\end{document}